\documentclass[sigconf]{acmart}
\AtBeginDocument{%
  }

\copyrightyear{2025}
\acmYear{2025}
\setcopyright{cc}
\setcctype[4.0]{by-sa}
\acmDOI{10.1145/3772052.3772236}
\acmConference[SoCC '25]{ACM Symposium on Cloud Computing}{November 19--21, 2025}{Online, USA}
\acmBooktitle{ACM Symposium on Cloud Computing (SoCC '25), November 19--21, 2025, Online, USA}
\acmISBN{979-8-4007-2276-9/25/11}
\renewcommand\footnotetextcopyrightpermission[1]{}

\acmSubmissionID{171}

\usepackage{amsmath,amsfonts}
\usepackage[ruled,vlined,linesnumbered]{algorithm2e}
\usepackage{algpseudocode}
\usepackage{balance}
\usepackage{url}
\usepackage{hyperref}
\usepackage{cleveref}
\crefname{section}{§}{§§}
\Crefname{section}{§}{§§}
\usepackage{graphicx}
\usepackage{makecell}
\usepackage{subcaption}
\usepackage{textcomp}
\usepackage{xcolor}
\usepackage{array}
\usepackage{tabularx} 
\usepackage{booktabs} 
\usepackage{multirow}
\usepackage{todonotes}
\usepackage{tikz}
\usepackage{float}

\newcommand{\proj}{{\sc 10Cache}}

\newcommand{\evict}[1]{%
  \tikz[baseline=(char.base)]{%
    \node[shape=circle, draw=none, fill=brown, text=white, inner sep=1pt, 
          minimum size=1.1em, font=\footnotesize\bfseries] (char) {#1};}%
}

\newcommand{\prefetch}[1]{%
  \tikz[baseline=(char.base)]{%
    \node[shape=circle, draw=none, fill=green, text=black, inner sep=1pt, 
          minimum size=1.1em, font=\footnotesize\bfseries] (char) {#1};}%
}

\begin{document}

\title{\proj: Heterogeneous Resource-Aware Tensor Caching and Migration for LLM Training}


\author{Sabiha Afroz}
\affiliation{%
  \institution{Virginia Tech, USA}
  \state{}
  \country{}
  }
\email{sabihaafroz@vt.edu}

\author{Redwan Ibne Seraj Khan}
\affiliation{%
  \institution{Virginia Tech, USA}
  \state{}
  \country{}
  }
  \email{redwan@vt.edu}

\author{Hadeel Albahar}
\affiliation{%
  \institution{Kuwait University, Kuwait}
  \state{}
  \country{}
  }
  \email{hadeel.albahar@ku.edu.kw}

\author{Jingoo Han}
\affiliation{%
  \institution{Virginia Tech, USA}
  \state{}
  \country{}
  }
  \email{jingoo@vt.edu}

\author{Ali R. Butt}
\affiliation{%
 \institution{Virginia Tech, USA}
 \state{}
 \country{}
 }
 \email{butta@cs.vt.edu}







\begin{abstract}

Training large language models (LLMs) in the cloud faces growing memory bottlenecks due to the limited capacity and high cost of GPUs. While GPU memory offloading to CPU and NVMe has made large-scale training more feasible, existing approaches suffer from high tensor migration latency and suboptimal device memory utilization, ultimately increasing training time and cloud costs. To address these challenges, we present \proj, a resource-aware tensor caching and migration system that accelerates LLM training by intelligently coordinating memory usage across GPU, CPU, and NVMe tiers. \proj~profiles tensor execution order to construct prefetch policies, allocates memory buffers in pinned memory based on tensor size distributions, and reuses memory buffers to minimize allocation overhead.

Designed for cloud-scale deployments, \proj~improves memory efficiency and reduces reliance on high-end GPUs. Across diverse LLM workloads, it achieves up to 2$\times$ speedup in training time, improves GPU cache hit rate by up to 86.6$\times$, and increases CPU/GPU memory utilization by up to 2.15$\times$ and 1.33$\times$, respectively, compared to state-of-the-art offloading methods. These results demonstrate that \proj~is a practical and scalable solution for optimizing LLM training throughput and resource efficiency in cloud environments.

\end{abstract}






\maketitle

\section{Introduction}\label{inroduction}

Transformer-based large language models (LLMs) have become foundational in natural language processing and code generation due to their ability to capture complex context. As their accuracy improves with scale, LLMs continue to grow in size, reaching hundreds of billions or even trillions of parameters. Training such models, e.g., LLaMA 3 (70B)~\cite{llama} or GPT-4 (1.76T)~\cite{gpt4_technical_report}, demands massive compute and memory, often involving hundreds or thousands of GPUs. For instance, the 175B GPT model requires approximately 326 GB in FP16 format, which far exceeds the 80 GB capacity of a single NVIDIA H100~\cite{nvidia_h100_gpu} GPU. These constraints have driven both industry and academia to seek training solutions that operate efficiently on cloud-scale infrastructure.

This explosive growth in model size has introduced new challenges for cloud systems. Public cloud platforms (e.g., AWS~\cite{aws, aws_trainium}, Azure~\cite{azure, azure_ai}, GCP~\cite{gcp}) and private AI clusters face mounting pressure to maximize resource utilization and reduce the cost per training job. GPU memory constraints are especially acute: while GPU computational throughput continues to improve, memory capacity has not kept pace. For example, the NVIDIA H100 delivers 2$\times$ more FLOPS than the A100, yet offers only a marginal increase in memory size. This widening compute-memory imbalance makes GPU memory a primary bottleneck for scaling LLM workloads in the cloud, driving the need for efficient memory management and offloading strategies.


Many cloud users aim to reduce training costs by fine-tuning pre-trained LLMs on domain-specific tasks~\cite{universal_language_model_finetuning, finetuning_language_model_finetuning_biomedical_natural_language}. Although fine-tuning requires fewer iterations and less data than full model training, it still consumes substantial GPU memory~\cite{smartinfinity-HPCA24}, often hitting the same memory wall~\cite{beyond_the_memory_wall} that limits full-scale training. Expanding to multiple GPUs is not always practical due to cost and resource constraints in cloud environments, making memory-efficient single-GPU fine-tuning a critical capability for workloads.




\begin{figure}[ht]
  \centering
  \vspace{-1em}\includegraphics[width=0.7\linewidth]{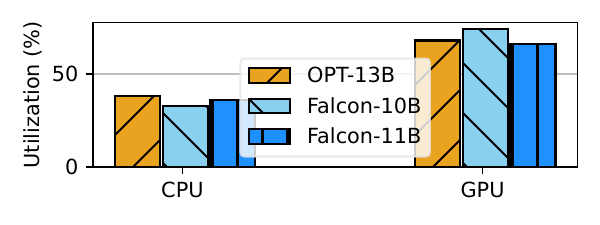}
  \vspace{-2em}
  \caption{CPU and GPU memory utilization in ZeRO-Infinity.}
  \Description{CPU and GPU memory utilization in ZeRO-Infinity.}
  \label{fig:cpu_gpu_mem}
  \vspace{-1em}
\end{figure}


To address memory bottlenecks in LLM training, especially in single GPU and cost-sensitive cloud environments, a range of techniques have been proposed, including mixed-precision arithmetic~\cite{mixed_precision_training, proteus, deep_learning_numerical_precision}, data compression~\cite{buddy_compression, gist}, and memory offloading to CPU and NVMe storage~\cite{autotm, swapadvisor, sentinel, flashneuron-FAST21, deepum-ASPLOS23, g10-MICRO23, zero_offload-ATC21, zero_infinity-SC21, l2l}. Among these, memory offloading is a practical and widely adopted strategy to enable GPU memory oversubscription. However, offloading tensors to CPU or NVMe adds high data migration latency, increasing training time, and reducing hardware efficiency, which is detrimental in cloud-scale deployments.

Mitigating this overhead requires an efficient and latency-aware memory migration strategy. A well-optimized offloading mechanism must not only expand usable memory capacity but also preserve GPU throughput by minimizing data transfer times. Our key observation is that existing solutions fail to fully utilize the available system memory hierarchy, resulting in underutilization of both CPU and GPU during training. Thus, improving memory migration efficiency and resource utilization is essential for enabling fast, scalable, and cost-efficient LLM training in the cloud.


To better understand these challenges, we perform a motivational study. 
We train, by fine-tuning, three models (OPT-13B, Falcon-10B, and Falcon-11B) for one epoch using DeepSpeed ZeRO-Infinity~\cite{zero_infinity-SC21} and observe their CPU and GPU memory utilization. ZeRO-Infinity offloads memory by partitioning model parameters, gradients, and optimizer states across CPU and NVMe, allowing larger models to fit within limited GPU memory. However, as shown in Fig.~\ref{fig:cpu_gpu_mem}, ZeRO-Infinity achieves suboptimal CPU and GPU memory utilization, ranging between 38\% and 74\%, due to inefficient memory offload during training. This inefficiency in resource utilization directly impacts training performance, leaving significant room for improvement. These findings underscore the need for an offloading strategy that maximizes the usage of available CPU and GPU computational power while minimizing tensor migration latency to enhance training efficiency.

Recognizing these challenges, we introduce \proj, a lightweight and resource-aware tensor caching and migration framework that improves memory efficiency and training throughput in multi-tier memory systems. \proj~targets three tiers of memory: GPU, CPU, and NVMe storage. It leverages a lightweight profiling phase to analyze tensor execution order, usage frequency, and size distribution. Based on this analysis, it constructs a prefetch table to proactively stage tensors in the fastest available memory (Table~\ref{tab:data_transfer_time}) before they are needed. To reduce the overhead of frequent allocations, \proj~pre-allocates pinned memory buffers and reuses them across training iterations, eliminating repeated allocation costs while improving data transfer performance via direct memory access.


By intelligently aligning tensor placement with access patterns, \proj~substantially reduces swap-in latency, boosts cache hit rates, and increases GPU and CPU memory utilization during training. These improvements allow large LLMs to be trained more efficiently using fewer or lower-cost GPUs by effectively leveraging underutilized CPU and NVMe memory tiers, making \proj~particularly valuable for cost-sensitive and resource-constrained cloud environments, where maximizing hardware efficiency is critical. In practical fine-tuning scenarios, \proj~improves training throughput while minimizing reliance on expensive multi-GPU setups, ultimately reducing operational costs and power consumption~\cite{cloud_llm_power}. These capabilities offer tangible economic and environmental benefits for cloud providers and AI infrastructure operators.
Overall, the major contributions of this paper are summarized as follows:


\begin{itemize}
\item 
We analyze tensor execution order in deep learning workloads and propose a prefetching technique specifically designed for LLM training, improving training efficiency.

\item 
We introduce a dynamic hierarchical tensor allocation mechanism that distributes tensors across GPU, CPU, and NVMe memory, increasing the GPU cache hit rate.

\item 
We design a novel resource-aware pre-allocated cache buffer for both CPU and GPU, reducing memory allocation overhead and enabling efficient memory management during tensor migration.


\item
We integrate \proj~into the widely-used DeepSpeed framework~\cite{deepspeed} and compare it against eight state-of-the-art baseline approaches. Our evaluations show that \proj~reduces the number of tensors with wait time below 0.03~ms by up to~$1.92\times$, increases the GPU cache hit rate by up to 86.6$\times$, CPU memory utilization by up to 2.15$\times$, GPU memory utilization by up to 1.33$\times$, and thus reduces training time by up to 2$\times$ compared to state-of-the-art methods. 


\end{itemize}

Our results demonstrate that \proj~is a practical, deployable system for improving LLM training efficiency in heterogeneous, memory-constrained environments, offering immediate impact for cloud-scale training platforms. \proj~is publicly available at https://github.com/Sabiha1225/10cache.git



\begin{table}[htbp]
\centering
\caption{Data transfer bandwidths across system components.}
\label{tab:data_transfer_time}
\vspace{-1em}
\begin{tabular}{|c|c||c|c|}
\hline
 Transfer Type & Bandwidth & Transfer Type & Bandwidth \\ 
 \hline
 CPU-GPU & 10.36 GB/s & CPU-NVMe Write & 0.73 GB/s \\
 \hline
 GPU-CPU & 9.51 GB/s & CPU-NVMe Read & 2.36 GB/s \\
 \hline
\end{tabular}
\vspace{-1em}
\end{table}

\section{Background and motivation}\label{background}

\subsection{Memory Demands in LLM Training}\label{llm_memory_requirement}

\begin{figure}
  \centering
\vspace{-1em}  \includegraphics[width=0.8\linewidth]{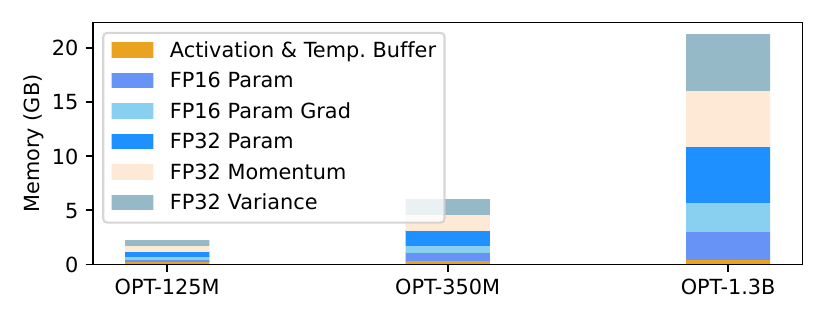}
  \vspace{-1.8em}
  \caption{LLM model states memory breakdown.
  }
  \Description{Memory requirements for LLM parameters, gradients, and optimizer states.}
  \vspace{-1.3em}
  \label{fig:model_memory_requirement}
\end{figure}

Training a DNN consists of three key steps: (1) forward pass, (2) backward pass, and (3) parameter update. In LLMs, these steps demand substantial memory, primarily due to model states, which include parameters, gradients, and optimizer states (e.g., momentum and variance in the Adam optimizer~\cite{adam}) required for mixed-precision training (FP16/32)~\cite{mixed_precision_training}. The remaining memory is consumed by activations and temporary buffer~\cite{zero-SC20}. Mixed-precision training~\cite{mixed_precision_training} with NVIDIA GPUs improves tensor core utilization~\cite{nvidia_tesla_architecture, zero-SC20} by running forward and backward passes in FP16, storing parameters and activations in FP16 format.
However, during the parameter update step with the Adam optimizer, large models require extra memory for FP32 copies of parameters, momentum, variance, and gradients. Specifically, for a model with $N$ parameters, FP16 copies of parameters and gradients require $2N$ bytes each, while FP32 copies of parameters, momentum, and variance each require $4N$ bytes~\cite{zero-SC20, zero_infinity-SC21}. Fig.~\ref{fig:model_memory_requirement} shows the memory breakdown for OPT-125M, OPT-350M, and OPT-1.3B, revealing that model states consume significantly more memory than activations and buffers.

Many previous works~\cite{zero_offload-ATC21, l2l} have addressed the GPU memory wall by offloading optimizer states to CPU memory and harnessing CPU computation for parameter updates when training large Transformer-based models. However, these approaches have limitations. ZeRO-Offload~\cite{zero_offload-ATC21} stores all model parameters in GPU, making GPU memory the limiting factor for training large models. In contrast, L2L~\cite{l2l} keeps only the current execution layer in GPU, leading to poor GPU memory utilization. For billion and trillion parameter models~\cite{llama, gpt4_technical_report}, neither GPU nor CPU memory alone is sufficient. To address this, some works~\cite{zero_infinity-SC21, optimstore} use NVMe storage offloading. ZeRO-Infinity enables offloading of both parameters and optimizer states to CPU and NVMe. However, existing offloading strategies often fail to utilize GPU memory efficiently when running large models, resulting in suboptimal memory usage. An effective solution should dynamically load model parameters to maximize GPU memory utilization without making the GPU a performance bottleneck. To this end, if the GPU retains only the tensors immediately needed for computation, up to its memory capacity, it can ensure timely access to minimize offloading-induced delays.

\subsection{Impact of Pinned vs. Pageable Memory on CPU–GPU Transfer Efficiency} \label{pinned_memory}

GPU memory offloading involves several data transfers between CPU and GPU, affecting LLM training time. CPU memory allocation uses two types: pageable and pinned memory. In pageable memory, the GPU cannot access data directly; the CUDA driver first creates a temporary pinned (page-locked) array~\cite{pinned_host_memory}, copies data from pageable memory to it, and then transfers it to GPU. 
Pinned memory serves as a staging area for transfers from GPU to CPU.


Our experimental results (Table~\ref{tab:pageable_pinned_memory_transfer}) show that transferring data from CPU-pinned memory to GPU takes less than half the time compared to pageable CPU memory. This observation motivates the use of pinned cache memory in our proposed \proj~system to reduce data transfer time during LLM training. Although DeepSpeed~\cite{deepspeed} uses pinned memory to store tensors, it keeps them in the same memory space throughout training. In contrast, \proj~takes advantage of faster data transfers to and from the GPU by reusing pre-allocated pinned memory. Although pinned memory takes longer to allocate than pageable memory, \proj~performs this allocation offline, so it does not add any overhead during model training. Additionally, frequent memory allocations in offloaded training increase training time, whereas reusing pre-allocated memory offers a more efficient solution. \proj~leverages these insights for effective memory management.

\begin{table}[h]
\centering
\vspace{-0.8em}
\caption{Data transfer time and bandwidth comparison: Pageable vs. Pinned memory.}
\label{tab:pageable_pinned_memory_transfer}
\renewcommand{\arraystretch}{1.1}
\setlength{\tabcolsep}{0.2pt} 
\vspace{-1em}
\begin{tabular}{|c|c|c|c|c|c|c|}
\hline
\textbf{Type} & \textbf{CPU-GPU} & \textbf{CPU-GPU} & \textbf{GPU-CPU} & \textbf{GPU-CPU} & \textbf{Data} & \textbf{Size} \\  
 & \textbf{(ms)} & \textbf{(GB/s)} & \textbf{(ms)} & \textbf{(GB/s)} & \textbf{Type} & \textbf{(MB)} \\  
\hline
Pageable  & 1.65 & 10.16 & 1.68 & 10.00 & FP32 & 16 \\  
\hline
Pinned    & 0.68 & 24.74 & 0.65 & 25.91 & FP32 & 16 \\  
\hline
Pageable  & 0.78 & 10.69 & 0.89 & 9.48  & FP16 & 8  \\  
\hline
Pinned    & 0.34 & 24.44 & 0.33 & 25.70 & FP16 & 8  \\  
\hline
\end{tabular}
\end{table}



\subsection{Tensor Behavior Analysis}\label{tensor_behavior_analysis}
A better understanding of tensor behavior in LLM training can help optimize GPU memory offloading and faster training. To study tensor execution patterns, active-inactive periods, and usage frequency, we generated the OPT-125M model trace using PyTorch's FX graph~\cite{fx_graph, pytorch}. The analysis reveals a tensor's life cycle: it is created during a PyTorch operation, used once or multiple times, and garbage collected when no longer needed.

\begin{figure}
  \centering
  \includegraphics[width=1.0\linewidth]{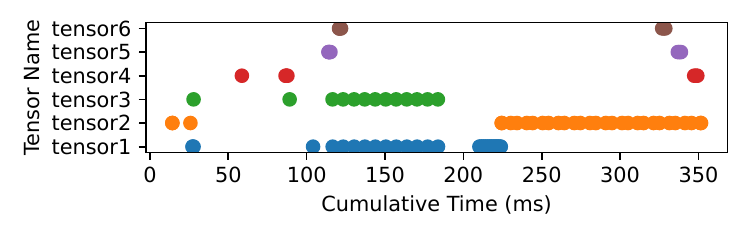}  
  
  \vspace{-1.5em}
  \caption{Tensors timeline for the OPT-125M model. Each dot marks a unique PyTorch operation at which the corresponding tensor becomes active.}
  \Description{Tensor timeline for OPT-125M model.}
  \label{fig:tensor_timeline}
\end{figure}

From the OPT-125M FX graphs, we computed tensors' active and inactive times, usage frequency, and observed tensor size variations across layers. Fig.~\ref{fig:tensor_timeline} shows the six most frequently used tensors and their timelines. Although created in different layers, these tensors are reused across multiple PyTorch operations within a single training iteration. The x-axis shows the cumulative execution time (ms), and the y-axis lists tensor names. Fig.~\ref{fig:tensor_timeline} shows that ``tensor1'' is used at different times in multiple operations. This repeated usage pattern underscores the importance of tensor caching.



Active time indicates when a tensor participates in a PyTorch operation, while inactive time denotes periods of idleness. For example, in Fig.~\ref{fig:tensor_timeline}, ``tensor2'' becomes active at 15~ms (1st torch operation), remains active for 10~ms (till 2nd torch operation), then stays idle for 200~ms before becoming active again at 225~ms (3rd torch operation). During the 200~ms idle window, the tensor can be offloaded to CPU or NVMe memory and fetched back to GPU memory when needed, improving memory efficiency.



According to the FX graphs, about 90\% of the kernels in the OPT-125M, OPT-1.3B, and OPT-2.7B models complete within 0.10~ms, 0.11~ms, and 0.12~ms, respectively, while about 50\% finish within 0.09~ms. This motivates our choice of a 0.03~ms threshold for tensor wait time analysis (\cref{wait_time_cpu_gpu},~\cref{wait_time_cpu_gpu_nvme}), as it remains well below typical kernel durations, minimizing the impact of swap-in latency on GPU throughput. 
To study system behavior across a range of tolerances, we also vary the threshold values (e.g., 0.01~ms, 0.1~ms) (\cref{wait_time_cpu_gpu}).



In this work, we focus on LLM training workloads, which generally operate with a static execution graph and exhibit predictable repeated tensor operation patterns during training~\cite{deepum-ASPLOS23, g10-MICRO23}. Our FX-graph analysis further confirms this regularity.
\proj~is designed to exploit these repeated patterns to allow efficient tensor caching and smart prefetching and eviction.
As a result, models with dynamic execution graphs, where tensor behavior is irregular and harder to predict, are beyond the scope of this study.

Training an LLM involves forward pass, backward pass, and parameter updates, requiring efficient memory management across GPU, CPU, and NVMe. While ZeRO-Infinity~\cite{zero_infinity-SC21} stores larger tensors in CPU or NVMe memory, \proj~keeps them in GPU memory if they are immediately required. G10~\cite{g10-MICRO23} considers the available bandwidth of flash storage and host memory when offloading tensors. Ideally, larger tensors should reside in GPU memory when needed to ensure immediate availability and avoid delays.

\section{\proj~Design}\label{design}


\begin{figure*}[t]
  \centering
\includegraphics[width=0.7\linewidth]{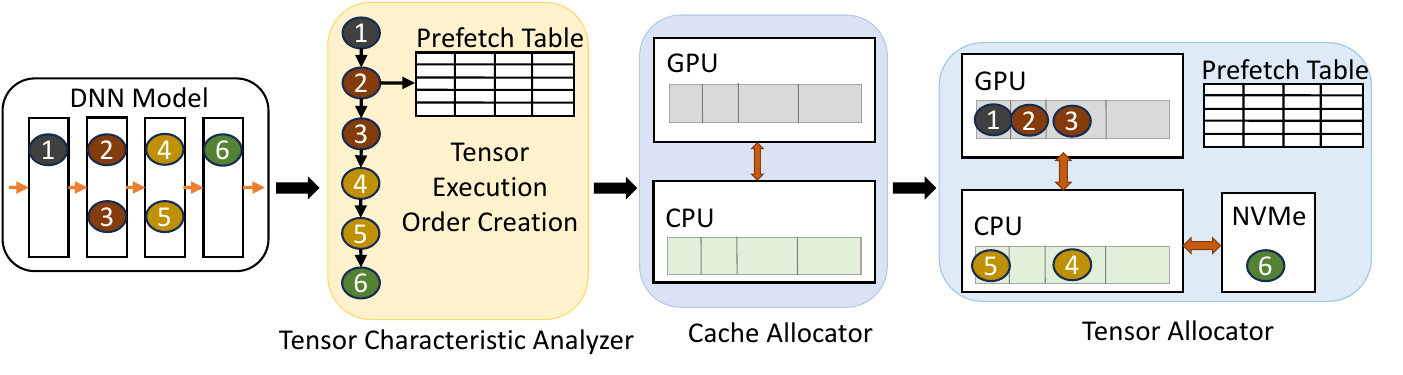}
  \vspace{-1em}
  \vspace{-0.4em}
  \caption{\proj's three components' end-to-end flow from profiling a DNN model to tensor allocation across CPU, GPU and NVMe memory. Tensors from the same layer share the same color, while different layers use distinct colors.} 
  \vspace{-1em}
  \label{fig:full_system_architecture_a}
\end{figure*}





\proj~is an efficient training framework designed to scale large models on a single GPU, using heterogeneous memory tiers, including CPU and NVMe storage, to overcome GPU memory limits and improve training throughput over baselines~\cite{zero_infinity-SC21, stronghold-SC22}. Its fine-grained, resource-aware tensor placement and dynamic prefetch-eviction strategies ensure optimal GPU memory usage. \proj~seamlessly integrates into existing workflows, making it both powerful and user-friendly. This section outlines the \proj~design.

\subsection{System Overview}

\proj~consists of four key components: the tensor characteristic analyzer, cache allocator, tensor allocator, and prefetch-eviction scheduler. Fig.~\ref{fig:full_system_architecture_a} shows the workflow of the first three components. The tensor characteristic analyzer extracts tensor execution order and size from the model to understand tensor behavior. Using this information, the cache allocator assigns cache buffers across GPU and CPU memory to maximize memory efficiency, while the tensor allocator places tensors to minimize unnecessary offloading. During training, the scheduler asynchronously prefetches and evicts tensors, overlapping data transfers with GPU computation (Fig.~\ref{fig:full_system_architecture_b}, Fig.~\ref{fig:full_system_architecture_c}). The following sections detail each component.

\subsection{Tensor Characteristic Analyzer}

To efficiently utilize limited and costly GPU and CPU memory, \proj~applies intelligent tensor migration policies that exploit heterogeneous memory while minimizing offloading overhead. This requires understanding tensor behavior and memory demands. The tensor characteristic analyzer performs a dry run of the model to capture execution patterns and tensor sizes. Based on this analysis, \proj~builds a prefetch table and characterizes tensor sizes.


\subsubsection{Prefetch Table Creation} \label{prefetch_table_creation}

When utilizing multi-tiered storage to offload tensors for large-model training, the tensor prefetch-eviction scheduler must anticipate when a tensor will become active in a GPU kernel operation. Without this foresight, delayed tensor retrieval can lead to GPU stalls and prolonged training times. To enable efficient prefetching and eviction, \proj~builds a prefetch table that records each tensor's execution order, activation time, current location, and final location. The current location indicates where the tensor resides during training (CPU, GPU, or NVMe), while the final location refers to the optimal memory tier assigned by the tensor allocator before training begins. After each iteration, tensors are restored to their final locations to maintain optimal placement. PyTorch~\cite{pytorch_hook} allows registering \texttt{pre\_hook} and \texttt{post\_hook} functions that run before and after a layer executes during forward and backward passes. \proj~employs these hooks to track tensor execution patterns, 
order and activation times. From this analysis, it builds a prefetch table which is later used for tensor placement (\cref{tensor_allocator}) in multi-tiered memory and for prefetch-eviction of tensors during training (\cref{training}). \proj~utilizes the observation that the tensor execution order remains consistent and repetitive throughout the DNN training~\cite{deepum-ASPLOS23} (\cref{tensor_behavior_analysis}). Therefore, capturing and using this execution order for tensor prefetching and eviction during training becomes both effective and desirable.  



\subsubsection{Tensor Size Characterization}

To manage memory efficiently and reduce the frequent allocations overhead during tensor offloading, \proj~pre-allocates dedicated memory buffers for tensor caching. A key challenge in pre-allocation is the variation in tensor sizes within a model. A uniform memory allocation would lead to internal and external fragmentation~\cite{memo, gmlake}, wasting valuable GPU and CPU memory critical for LLM training. To address this, the tensor characteristic analyzer performs a lightweight dry run to profile and categorize tensor sizes, incurring minimal overhead compared to total training time (\cref{profiling_overhead}). Based on this, \proj~optimally allocates cache memory between GPU and CPU, reducing fragmentation and maximizing memory efficiency. \proj~organizes buffers by tensor sizes to ensure seamless tensor caching and retrieval during training (\cref{buffered_memory_allocation}).

\subsection{Cache Allocator}

As analyzed in~\cref{pinned_memory}, tensor swap-in and swap-out memory operations are costly. To reduce this overhead, \proj~adopts two key strategies. First, it performs memory allocation once and reuses the same memory for all subsequent swap-in and swap-out operations. This eliminates the time spent on repeated allocations during training, leading to improved training time (\cref{cpu_gpu_training_time}). An additional advantage of reusing memory is that it reduces fragmentation, which often results from frequent allocations during training. This design choice not only speeds up training, but also ensures efficient memory utilization. Second, \proj~strategically allocates memory in pinned CPU space to speed up CPU–GPU data transfers. Although pinned memory allocation takes longer than pageable memory, \proj~performs this allocation once before training begins, so it does not impact the overall training time. 

\subsubsection{Buffered Memory Allocation} \label{buffered_memory_allocation}
A key challenge lies in allocating memory that can be reused without creating internal or external fragmentation. As observed in~\cref{tensor_behavior_analysis}, tensor sizes vary across different model layers. While one might consider allocating all buffers to match the size of the largest tensor in both CPU and GPU, this naive approach leads to internal memory fragmentation. For example, when storing a 512-byte tensor in a 1024-byte buffer (sized for the largest tensor), half of the buffer space goes unused. A more efficient approach analyzes the specific tensor size distribution of each model, allowing optimized buffer allocation that minimizes memory waste. The pre-allocation process involves three steps: 1) calculating the tensor size distribution, 2) determining the required buffer count per size, and 3) allocating memory accordingly.

\begin{algorithm}[t!]
\caption{Tensor Size Distribution Calculator}\label{alg:tensor_size_distribution_calculator}
\KwData{$TC = \{(s_1, c_1), (s_2, c_2), \dots\} $ \tcp*{TC = a dict for tensor count $(c_i)$ for each size $(s_i)$} } 
\KwResult{$TSD = \{\}$  \tcp*{TSD = a dict holding the distribution for each size}} 
$\text{total\_size} \gets 0$\;
\ForEach{$(s_i, c_i) \in TC$}{
    $sc \gets s_i * c_i$\;
    $\text{total\_size} \gets \text{total\_size} + sc$\;
    $TSD[s_i] \gets sc$\;
}
\ForEach{$s_i \in TSD$}{
    $TSD[s_i] \gets TSD[s_i] / total\_size$\;
}
\end{algorithm}


\textbf{Step 1: Tensor Size Distribution Calculation.} This process takes as input a dictionary $TC$, where each entry $(s_i, c_i)$ represents the tensor size ($s_i$) and its corresponding count ($c_i$). It then computes the total memory requirement in bytes for each unique tensor size (Alg.~\ref{alg:tensor_size_distribution_calculator}, lines 2–5). Subsequently, it calculates the memory requirement ratio of each tensor size relative to the total memory requirement for all tensors (Alg.~\ref{alg:tensor_size_distribution_calculator}, lines 6–7). 



\textbf{Step 2: Buffer Count by Tensor Size in Cache Memory.} This process computes the number of buffers for each tensor size based on available GPU and CPU memory. It takes as input the tensor size distribution from Step~1 and uses two memory profiler APIs
to obtain the available GPU and CPU memory (Alg.~\ref{alg:buffer_counter}, lines~1-2). Using these data, the buffer counter computes the required buffers per tensor size, first for FP16 parameters on GPU (Alg.~\ref{alg:buffer_counter}, line~4), then for FP16 tensors on CPU (Alg.~\ref{alg:buffer_counter}, line~5), and finally for FP32 optimizer states.

\textbf{Step 3: Memory Allocation.} This step utilizes the buffer counts computed in Step 2 (GPU\_BUFFER\_COUNT, CPU\_BUFFER\_COUNT) to determine the total memory required for each tensor type. To reduce fragmentation, \proj~allocates a contiguous region in pinned memory and partitions it into fixed-size chunks based on buffer sizes. A free buffer list tracks available buffers for reusability.



\begin{figure}[!t]
  \centering
  \includegraphics[width=\linewidth]{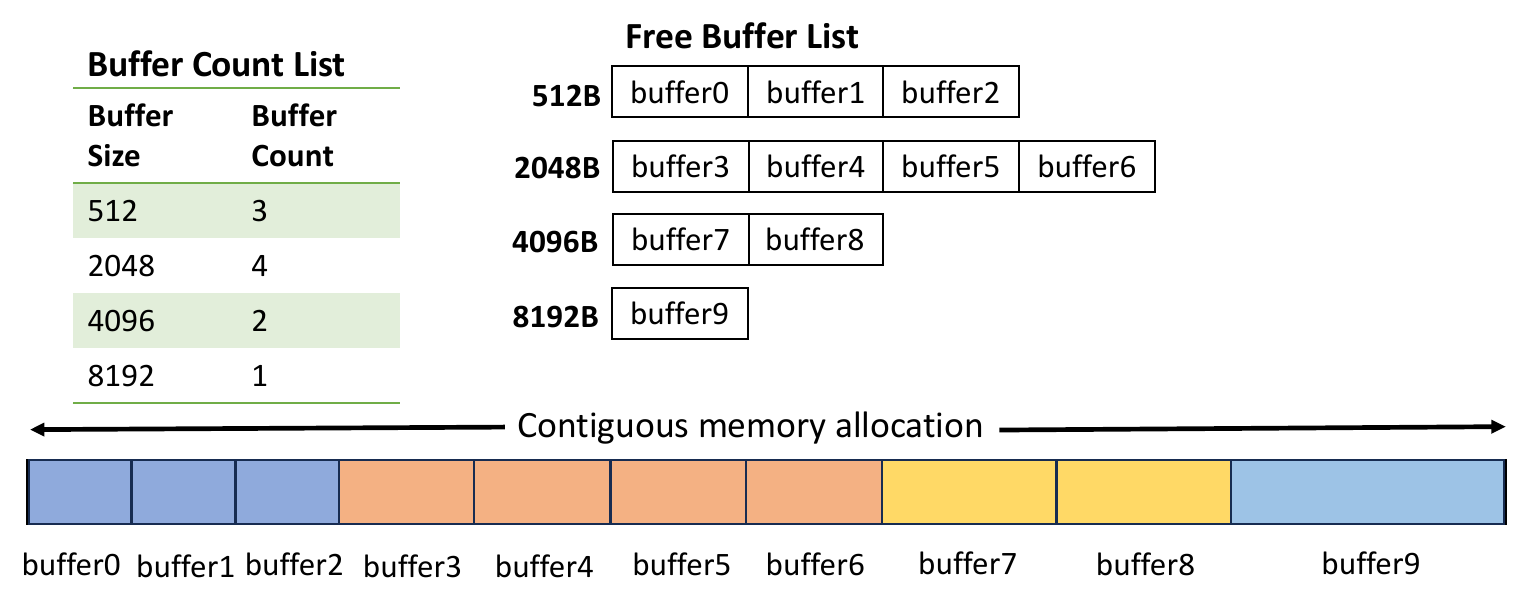}
  \vspace{-1.8em}
  \vspace{-0.4em}
  \caption{Cache buffer allocation strategy (A contiguous memory block is partitioned into fixed-size buffers for different tensor sizes. Buffers are registered in a free list by size.)}
  \vspace{-1em}
  \vspace{-0.4em}
  \label{fig:cache_buffer_allocation}
\end{figure}

From Fig.~\ref{fig:cache_buffer_allocation}, the table outlines the number of buffers required for each tensor size. Below it, the diagram represents a contiguous memory block. The allocation process begins by assigning the first 512-byte chunk as `buffer0', which is then added to the free buffer list. This process continues sequentially, allocating memory chunks
and registering them in the free buffer list according to their sizes. When the system requires a 512-byte buffer, it first retrieves `buffer0' from the free list, ensuring efficient memory utilization while reducing allocation overhead.

\begin{algorithm}[t]
\caption{Buffer Counter}\label{alg:buffer_counter}
\KwData{$TC = \{(s_1, c_1), (s_2, c_2), \dots\} $ , $TSD$ \tcp*{TC = a dict for tensor count $(c_i)$ for each size $(s_i)$, TSD = a dict holding the distribution for each size}}
\KwResult{$\text{GPU\_BUFFER\_COUNT} \gets \{\}$, $\text{CPU\_BUFFER\_COUNT} \gets \{\}$ \tcp*{two dict for holding buffer count for each size}}
$\text{gpu\_avl\_mem} \gets \text{memory\_profiler.get\_gpu\_free\_mem()}$\;
$\text{cpu\_avl\_mem} \gets \text{memory\_profiler.get\_cpu\_free\_mem()}$\;
\ForEach{$(s_i, c_i) \in TC$}{
    $\text{GPU\_BUFFER\_COUNT}[s_i] \gets \min((TSD[s_i] * \text{gpu\_avl\_mem})/s_i, c_i)$\;

    $\text{CPU\_BUFFER\_COUNT}[s_i] \gets \min((TSD[s_i] * \text{cpu\_avl\_mem})/s_i, c_i - \text{GPU\_BUFFER\_COUNT}[s_i])$\;
}
\end{algorithm}

\subsubsection{Tensor Allocator} \label{tensor_allocator}

Our proposed \proj~efficiently adapts to available memory resources. When GPU and CPU memory can accommodate model parameters and optimizer states, it confines storage to these high-speed memory tiers. However, when the model size exceeds the combined capacity of GPU and CPU memory, \proj~dynamically distributes parameters and optimizer states across GPU, CPU, and NVMe storage. This strategic placement enhances training efficiency and optimizes resource utilization, ensuring minimal offloading overhead and improved performance.

\begin{figure}[t]
  \centering
  \includegraphics[width=0.8\linewidth]{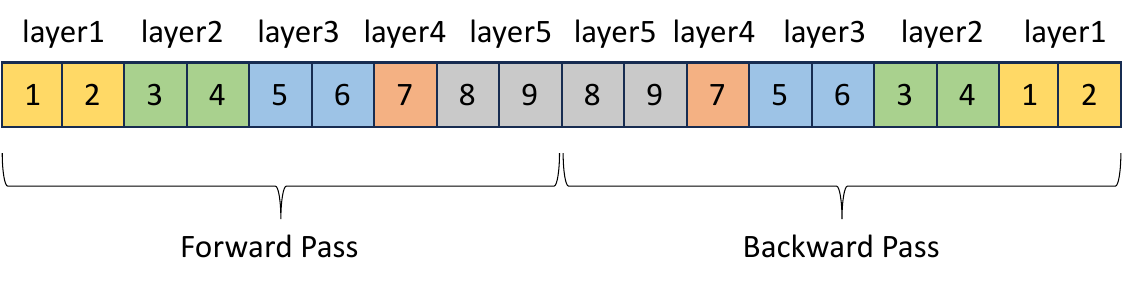}
   \vspace{-1em}
  \caption{A single training iteration for the model.}
  \vspace{-1em}
  \vspace{-5pt}
  \label{fig:model_one_iteration}
\end{figure}

\textbf{An Illustrative Example.} Fig.~\ref{fig:model_one_iteration} illustrates the execution sequence of FP16 parameters during the forward and backward passes in one
training iteration. The forward pass begins with layer 1, so tensors 1 and 2 from this layer are accessed immediately. If these tensors are large, ZeRO-Infinity~\cite{zero_infinity-SC21} stores them in CPU or NVMe memory, which introduces migration latency, as they must be brought back to GPU memory for computation. In contrast, \proj~leverages knowledge of tensor execution order to keep these immediately required tensors in GPU memory, significantly reducing migration latency. To optimize tensor placement, \proj's tensor allocator strategically distributes tensors across the tiered memory. It uses the prefetch table to identify tensors needed soon and prioritizes placing them in GPU cache. An active tensor window tracks tensors in GPU memory, while additional tensors are stored in CPU cache, with any remaining placed in NVMe if CPU memory is full.

In a scenario where CPU and GPU cache memory can accommodate the entire model, the \proj~tensor allocator ensures that tensors 1, 2, 3, and 4 remain in GPU memory, while tensors 5, 6, 7, 8, and 9 are stored in CPU. As the scheduler loads tensors 6, 7, 8, and 9 into GPU for execution, it retains them for
immediate reuse during backpropagation. This approach significantly increases the GPU cache hit rate (\cref{hit_rate_cpu_gpu} and~\cref{hit_rate_cpu_gpu_nvme}) and reduces the frequency of tensor offloading. The prefetch table maintains an up-to-date record of tensor locations across heterogeneous memory, enabling the scheduler to make informed prefetching and eviction decisions.

\subsection{Training}\label{training}

During model training, the \proj~scheduler efficiently manages tensor prefetching and eviction by overlapping GPU computation with CPU-GPU communication, thus minimizing data transfer overhead. For LLM training, the CPU optimizer~\cite{zero_offload-ATC21} performs parameter updates on the CPU, improving memory efficiency, since optimizer states consume substantial memory (\cref{llm_memory_requirement}) and GPU memory is both limited and expensive. Based on model size, \proj~dynamically employs either a CPU-GPU or an extended CPU-GPU-NVMe offloading strategy to scale large model training. To ensure robustness under runtime noise (e.g., OS jitter, multi-tenant contention) in production environments, \proj~avoids reliance on wall-clock timing. Instead, it uses lightweight PyTorch pre-/post-hooks (\cref{prefetch_table_creation}) that trigger prefetching based on actual layer execution, ensuring the scheduler remains synchronized even when execution drifts from the profiled timing.

\begin{algorithm}[t]
\DontPrintSemicolon
\caption{PrefetchTensor}\label{alg:prefetching}
\KwIn{evicted\_tensor\_list}

\ForEach{evict\_tensor\_id \textbf{in} evicted\_tensor\_list}{
    \If{prefetch\_tab\_cur\_row $<$ len(prefetch\_table)}{
        Remove evict\_tensor\_id from active\_tensor\_window\;
        \textsc{EvictTensor}(evict\_tensor\_id)\;
        \While{prefetch\_table[prefetch\_tab\_cur\_row].tensor\_id $\in$ active\_tensor\_window \textbf{and} prefetch\_tab\_cur\_row $<$ len(prefetch\_table)}{
            prefetch\_tab\_cur\_row++\;
            release\_param $\leftarrow$ \textbf{False}\;
        }
        prefetch\_tensor\_id $\leftarrow$ prefetch\_table[prefetch\_tab\_cur\_row].tensor\_id\;
        Add prefetch\_tensor\_id to active\_tensor\_window\;
        Get a free GPU buffer ID for prefetch\_tensor\_id\;
        \eIf{prefetch\_table[prefetch\_tab\_cur\_row].current\_loc is `cpu'}{
            Get CPU buffer ID for prefetch\_tensor\_id\;
            Async copy from CPU buffer to GPU buffer\;
        }{
            \If{prefetch\_table[prefetch\_tab\_cur\_row].current\_loc is `nvme'}{
                Async copy from NVMe to CPU temp buffer, then to GPU buffer\;
            }
        }
        prefetch\_tab\_cur\_row++\;
    }
}
\end{algorithm}


\begin{algorithm}[t]
\DontPrintSemicolon
\caption{EvictTensor}\label{alg:eviction}
\KwIn{evict\_tensor\_id}
Get final\_loc, evict\_tensor\_size, and GPU buffer ID for evict\_tensor\_id\;
\If{final\_loc is `nvme'}{
    Release GPU buffer for evict\_tensor\_id\;
    Update current location to `nvme'\;
}
\Else{
    \eIf{CPU has free buffer of size evict\_tensor\_size}{
        Get a free CPU buffer ID\;
    }{
        \eIf{CPU has occupied GPU-designated buffer of evict\_tensor\_size}{
            Get its tensor ID and buffer ID\;
            Swap tensor to NVMe, update location to `nvme'\;
            Release GPU-designated CPU buffer\;
        }{
            Get any occupied CPU buffer ID and tensor ID of evict\_tensor\_size\;
            Swap tensor to NVMe, update location to `nvme'\;
        }
    }
    \If{final\_loc is `gpu'}{
        Mark buffer as GPU-designated in CPU cache\;
    }
    Mark CPU buffer as occupied by evict\_tensor\_id\;
    Async copy from GPU to CPU buffer\;
    Release GPU buffer for evict\_tensor\_id\;
    Update current location to `cpu'\;
}

\end{algorithm}


\subsubsection{CPU-GPU Offloading}\label{CPU_GPU_architecture}

\begin{figure}[h]
  \centering
\vspace{-.5em}  \includegraphics[width=1.0\linewidth]{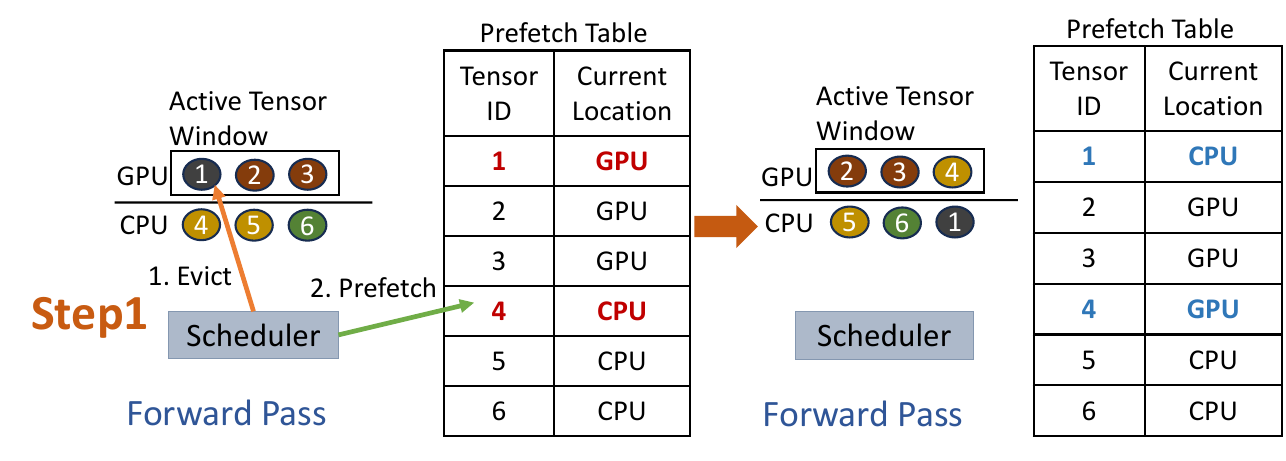}
  \vspace{-5pt}
  \vspace{-15pt}
  \caption{CPU-GPU Offloading: Prefetches tensor 4 and evicts tensor 1 during the forward pass based on execution order.}
  \label{fig:full_system_architecture_b}
\end{figure}
\vspace{-1.0em}

If the aggregate memory of the CPU and GPU is sufficient to store the model parameters and optimizer states, \proj~prevents storing any tensors in slower NVMe memory. In this scenario, the CPU cache stores all the optimizer tensors, while parameter tensors are distributed across the GPU and CPU cache. Fig.~\ref{fig:full_system_architecture_b}
demonstrates an example of tensor prefetching and eviction for CPU-GPU offloading. For example, the DNN model has parameter tensors 1, 2, 3, 4, 5, 6. \proj's tensor allocator places tensors 1, 2, and 3 in the GPU memory and tensors 4, 5, and 6 in CPU memory (Step1). During forward pass, once tensor 1 finishes execution, the scheduler evicts (\evict{1}) it to CPU (Alg.~\ref{alg:prefetching}, lines 1-4) and asynchronously prefetches (\prefetch{2}) tensor 4, which is not in active tensor window, ahead of time (Alg.~\ref{alg:prefetching}, lines 5-13). Therefore, when tensor 4 is needed for kernel execution, it will already be in the GPU. After eviction and prefetching, the GPU stores tensors 2, 3, 4, and the CPU stores tensors 1, 5, 6. Similarly, the scheduler evicts tensors 2 and 3 in CPU and prefetches tensors 5 and 6 in GPU during forward pass. When the activation window contains tensors 4, 5 and 6 in GPU, the scheduler halts any eviction or prefetching, as these tensors are essential for immediate backward pass. Offloading them would require reloading, incurring unnecessary overhead. By avoiding such redundant transfers, \proj’s scheduling policy improves both training time (\cref{cpu_gpu_training_time}) and cache hit rate (\cref{hit_rate_cpu_gpu}).

\subsubsection{CPU-GPU-NVMe Offloading}\label{CPU-GPU-NVMe Offloading}

\begin{figure*}[t]
  \centering
\vspace{-.5em}  \includegraphics[width=0.9\linewidth]{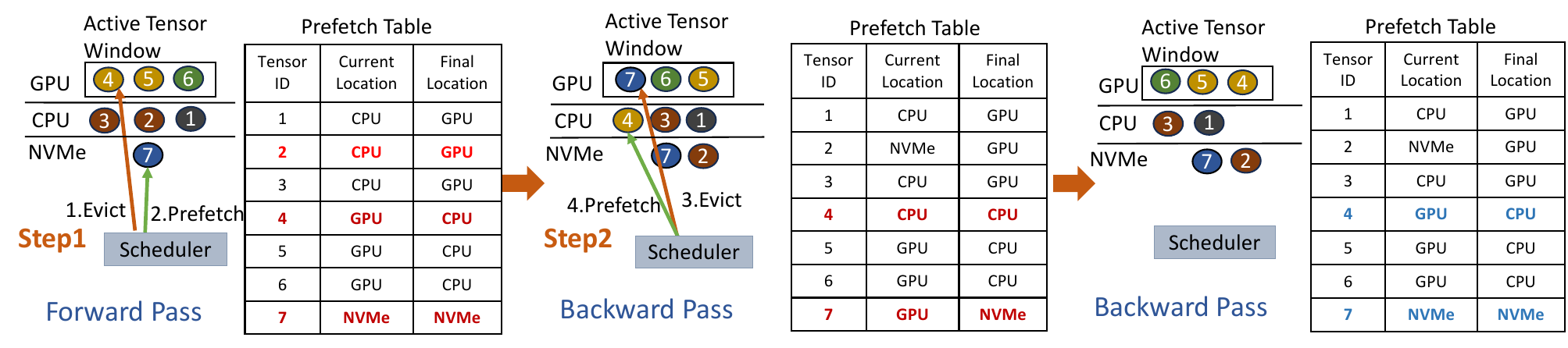}
  \vspace{-1em}
  \caption{CPU-GPU-NVMe offloading: when memory is limited, \proj's scheduler coordinates tensors across all three tiers.}
  \vspace{-1em}
  \label{fig:full_system_architecture_c}
\end{figure*}

For large models, CPU and GPU memory alone cannot accommodate all model parameters and optimizer states. In such cases, \proj's tensor allocator strategically stores a portion of the FP16 parameters and optimizer states in NVMe to scale LLM training. 





\textbf{FP16 Parameter Scheduling Policy.} 
Fig.~\ref{fig:full_system_architecture_c} demonstrates prefetching and eviction in a CPU-GPU-NVMe setup. In Step1, during the forward pass, tensors 1, 2, and 3 complete execution and \proj~offloads them to CPU cache. Tensors 4, 5, and 6 remain active in GPU memory, while tensor 7 is placed in NVMe by tensor allocator due to limited CPU and GPU memory. Once tensor 4 completes execution during the forward pass, the scheduler attempts to evict it to the CPU cache (\evict{1}). As the CPU cache lacks free buffer space and tensor 2 (currently in the CPU) has the same size as tensor 4 while remaining inactive for a longer period, \proj~evicts tensor 2 to NVMe storage to create space for tensor 4 as tensor 4 will be used earlier in the backward pass. Now, \proj's scheduler marks tensor 7 for prefetch (\prefetch{2}). \proj~first copies the tensor into a temporary buffer in CPU memory before asynchronously transferring it to the GPU buffer (Alg.~\ref{alg:prefetching}, lines 14-16). This ensures uninterrupted GPU computation. NVMe always holds tensor 7 copy. When needed, this tensor is prefetched to GPU from NVMe. For eviction, only the GPU memory buffer is released since the NVMe copy is already available.

If GPU buffers for a tensor size are exhausted, \proj~evicts a tensor not needed soon. If it already resides in NVMe, the GPU buffer is released (Alg.~\ref{alg:eviction}, lines 2-4). Otherwise, it checks for a free CPU buffer to offload the tensor and releases the GPU buffer (Alg.~\ref{alg:eviction}, lines 6-7, 16-21). If CPU buffer is unavailable, it evicts a GPU-designated CPU tensor to NVMe and reuses it (Alg.~\ref{alg:eviction}, lines 9-12, 16-21). A GPU-designated CPU tensor is one initially placed in GPU memory by the tensor allocator but later offloaded to CPU after its forward-pass 
completes. 
When the CPU buffer is full for this particular tensor size, \proj's scheduler selects a GPU-designated CPU tensor for eviction, as it will be needed later during backpropagation. The tensor allocator prioritizes placing tensors in GPU, then CPU, and finally NVMe storage. During backpropagation, tensors are used in reverse order, starting with those in NVMe, followed by CPU, and then GPU. \proj's scheduler is aware of this order and evicts tensors accordingly to minimize offloading overhead. If none is found, it evicts any occupied CPU buffer to NVMe and reuses the buffer (Alg.~\ref{alg:eviction}, lines 13-21). The tensor allocator sets each tensor’s location
in the prefetch table (\cref{tensor_allocator}).




After prefetching tensor 7 and evicting tensor 4, GPU stores tensors 5, 6 and 7, CPU holds tensors 1, 3 and 4 and NVMe stores tensors 2 and 7. \proj's scheduler now halts any eviction and prefetch operations as tensors 5, 6, 7 will be used immediately in backward pass. In Step2, once tensor 7 execution finishes during the backward pass, the scheduler evicts (\evict{3}) it. As NVMe already holds a copy of tensor 7, \proj~releases the GPU buffer of tensor 7 back to the free buffer list (Alg.~\ref{alg:eviction}, lines 2-4), allowing it to be reused later by another tensor of similar size. As tensor 4 needs to be processed during the backward pass, it is prefetched (\prefetch{4}) from the CPU cache using a swap-in operation to the GPU cache (Alg.~\ref{alg:prefetching}, lines 11-13). Tensor 4 has the same size as tensor 7 and hence reuses the free buffer of tensor 7 in the GPU cache which was evicted previously. After eviction and prefetching, GPU stores tensors 4, 5 and 6, CPU stores tensors 1 and 3 and NVMe stores tensors 2 and 7.

\textbf{Optimizer States Scheduling Policy.}
For LLMs, optimizer states consume substantial memory (\cref{llm_memory_requirement}), often exceeding available CPU capacity. A conventional offloading approach like ZeRO-Infinity~\cite{zero_infinity-SC21} stores all optimizer states in NVMe due to limited CPU memory, but this introduces inefficiencies. Assume that we have three FP32 parameter tensors: T1, T2, and T3. In ZeRO-Infinity, these tensors and their optimizer states reside in NVMe. During the optimizer step, the scheduler first loads T1 and its corresponding optimizer state into CPU memory. After updating T1, it moves the tensor back to NVMe and then fetches T2. This synchronous swap-in and swap-out process increases training time and leaves CPU memory underutilized.




\begin{figure}[h]
  \centering
\includegraphics[width=0.8\linewidth]{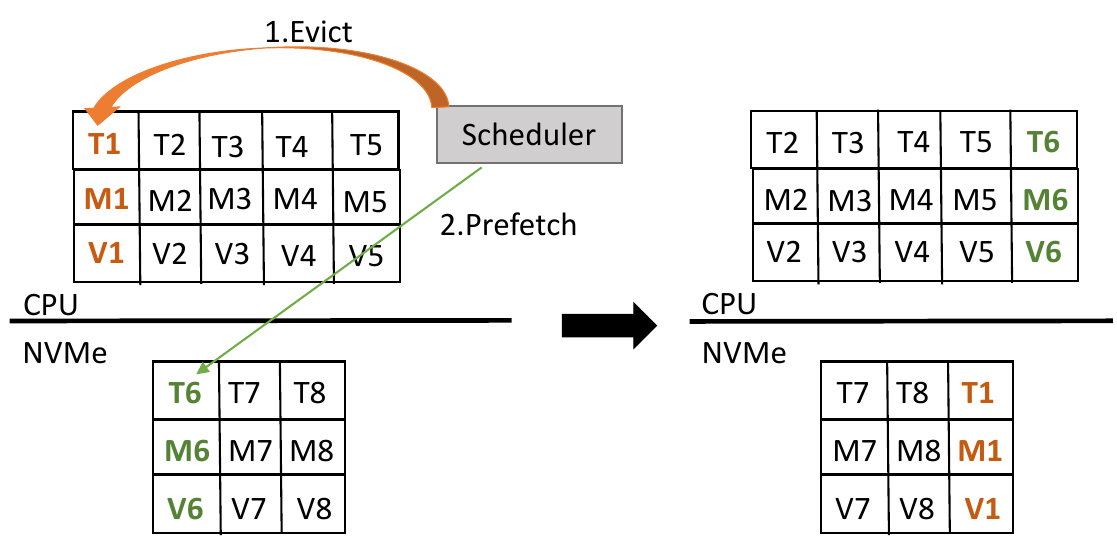}
  \vspace{-1em}
  \caption{Optimizer states prefetching and eviction.}
  \label{fig:fp32_prefetch_eviction}
\end{figure}

To overcome this inefficiency, \proj's tensor allocator dynamically distributes optimizer state tensors between CPU cache and NVMe based on available CPU memory capacity. As illustrated in Fig.~\ref{fig:fp32_prefetch_eviction}, \proj~stores optimizer states for tensors T1 through T5 in CPU, while T6, T7, and T8 are stored in NVMe. Once the parameter update for T1 is complete, the scheduler evicts it (\evict{1}) and asynchronously prefetches T6 (\prefetch{2}). This asynchronous prefetching overlaps data transfer time from NVMe to CPU with CPU computation. This strategy improves both training time (\cref{cpu_gpu_nvme_training_time}) and CPU memory utilization (\cref{cpu_gpu_memory_utilization}). 

\section{Evaluation}\label{evaluation}



\subsection{Experimental Setup}
\subsubsection{Implementation Details}



\proj~is built on top of DeepSpeed~\cite{deepspeed}, a PyTorch-based~\cite{pytorch} framework optimized for LLM training. It extends ZeRO-Infinity~\cite{zero_infinity-SC21} with efficient parameter and optimizer states placement across heterogeneous memory, intelligent tensor migration scheduling, and advanced memory management to enhance training throughput. \proj~integrates seamlessly with DeepSpeed, supporting any compatible LLM workflow for practical and efficient training. For profiling, it uses \texttt{nvidia-smi} to monitor GPU memory and \texttt{Psutil} to track CPU memory.


\subsubsection{System Configurations}
Table~\ref{tab:system_configuration} summarizes the experimental setup. In all configurations, the Samsung NVMe connects to the CPU via PCIe 4.0 ×4, and the GPU connects to the CPU via PCIe 4.0 ×16. We use the NVIDIA L40S GPU as the default, unless stated otherwise. All experiments use PyTorch 2.3.0 and DeepSpeed 0.14.2.



\begin{table}[h]
\centering
\caption{System configurations.}
\label{tab:system_configuration}
\vspace{-1em}
\begin{tabular}{ | p{0.12\linewidth}| p{0.38\linewidth}| p{0.38\linewidth} |} 
 \hline
 CPU & 2x Intel Xeon Silver 4314 (16c, 32t) & AMD EPYC 7763 64-Core Processor \\ 
 \hline
 GPU & NVIDIA L40S (48GB), A40 (48GB) & A100 (40GB) \\ 
 \hline
 Memory & 256GB, 200GB & 256GB \\ 
 \hline
 Storage & 2TB Samsung NVMe & 3TB Samsung NVMe \\ 
 \hline
\end{tabular}
\end{table}





\subsubsection{Workloads and Datasets}
We evaluate \proj~using diverse LLMs and datasets summarized in Table~\ref{tab:llm_models}. To emulate limited GPU memory scenarios, we use a default batch size of 16 and vary it in~\cref{batch_size_impact_on_training_time} to analyze its effect on training time.



\begin{table}[htbp]
\centering
\caption{LLM workloads and datasets.}
\label{tab:llm_models}
\vspace{-1em}
\begin{tabular}{|c|c|c|}
\hline
 Model & Source & Dataset \\ 
 \hline
 OPT-6.7B,OPT-13B~\cite{facebook-opt} & Hugging Face & GLUE MRPC\\
 \hline
 Bloom-7B~\cite{bloom} & Hugging Face & GLUE COLA\\
 \hline
 Falcon-7B,Falcon-10B & Hugging Face & Wikitext\\ 
 \hline
 Falcon-11B~\cite{falcon}, GPT2 & Hugging Face & Wikitext\\  
 \hline
\end{tabular}
\vspace{-1em}
\end{table}



\subsubsection{Baselines} \label{baselines}
We evaluate \proj~against eight recent state-of-the-art GPU memory swapping approaches. We ensure a fair evaluation by using the default or recommended configurations of baselines without extra prefetching or memory tuning beyond what these frameworks provide. 

ZeRO-Offload~\cite{zero_offload-ATC21} stores model parameters in GPU memory and optimizer states in CPU memory, relying on CPU computation for parameter updates. ZeRO-Infinity~\cite{zero_infinity-SC21} extends offloading across CPU, GPU, and NVMe. L2L~\cite{l2l} reduces GPU memory pressure by keeping only the active layer in GPU. StrongHold~\cite{stronghold-SC22} improves upon L2L by storing multiple layers to reduce offloading overhead. Megatron-LM~\cite{megatron-lm} enables large transformer training through model parallelism. FlashNeuron~\cite{flashneuron-FAST21} offloads intermediate tensors to NVMe SSD via direct GPU–SSD communication~\cite{gpudirect_storage}. DeepUM~\cite{deepum-ASPLOS23} uses CUDA Unified Memory~\cite{cuda_um} with a correlation-based prefetcher for GPU memory oversubscription. G10~\cite{g10-MICRO23} unifies host, GPU, and flash memory into a single memory space, scheduling tensor migration based on available bandwidth of flash and host memory.

For CPU-GPU offloading, we compare \proj~with ZeRO-Offload, ZeRO-Infinity, L2L, StrongHold, and Megatron-LM. For CPU-GPU-NVMe offloading, we evaluate it against ZeRO-Infinity, FlashNeuron, DeepUM, and G10. In CPU-GPU setups, \proj~denotes the base version without prefetching or memory optimizations. \proj+P adds prefetching using a prefetch table for timely tensor loading and eviction, and \proj+P+M adds pre-allocated buffers to reduce allocation overhead. For CPU-GPU-NVMe setups, \proj+FP16 applies all optimizations for FP16 parameters, while \proj+FP16+Opt further includes prefetching and eviction of optimizer states.

\subsection{Performance Analysis}

\subsubsection{Training Time Evaluation (CPU-GPU Offloading)} \label{cpu_gpu_training_time}

\begin{figure}[!htbp]
  \centering
\vspace{-1em}  \includegraphics[width=0.7\linewidth]{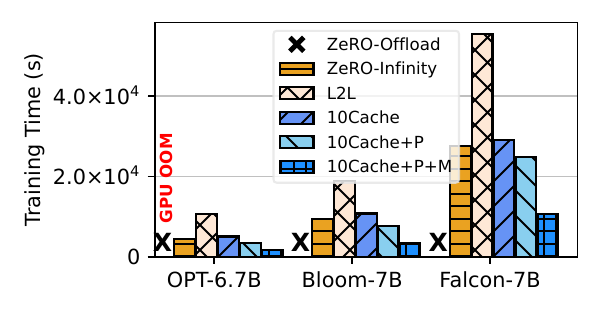}
  \vspace{-1em}
  \vspace{-1em}
  \caption{Training performance of \proj~vs. baselines under CPU-GPU offloading.}
  \vspace{-1em}
  \label{fig:training_time_cpu_gpu_6b_range}
\end{figure}

Fig.~\ref{fig:training_time_cpu_gpu_6b_range} compares the training times of OPT-6.7B, Bloom-7B, and Falcon-7B, showing how \proj~improves LLM training efficiency over baselines. In this experiment, we restrict GPU memory to 24 GB. ZeRO-Offload fails with GPU OOM error for 7B range models since it stores parameters in GPU memory and optimizer states in CPU memory (\cref{baselines}), requiring about 28 GB GPU memory for FP16 parameters and gradients, despite sufficient CPU RAM. ZeRO-Infinity supports parameter offloading to CPU memory, allowing training under this constraint. \proj+P+M reduces training time by about 2$\times$ compared with ZeRO-Infinity for all three models, while the base \proj~performs worse due to the lack of prefetching and pre-allocated cache memory. L2L performs the worst, loading one layer at a time and offloading it after execution, causing high communication overhead and GPU stalls.

\begin{figure}[!htbp]
    \centering
    
    \vspace{-1em}
    \begin{subfigure}[b]{0.40\textwidth}
        \centering
        \includegraphics[width=0.8\textwidth]{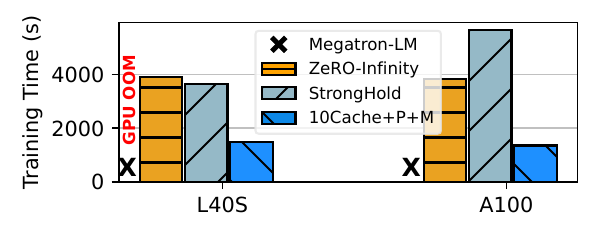}
         
        \vspace{-1em}
        \caption{Training efficiency comparison}
        \label{fig:training_time_cpu_gpu_gpt_5_9}
    \end{subfigure}\hfill
    \begin{subfigure}[b]{0.40\textwidth}
        \centering
        \includegraphics[width=0.8\textwidth]{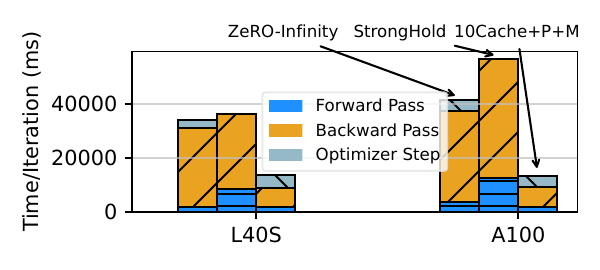}
        \vspace{-1em}
        \caption{Per-iteration time breakdown}
        \label{fig:one_iteration_execution_time_breakdown_gpt_5_9_b}
    \end{subfigure}
    \vspace{-0.6em}
    \vspace{-0.6em}
    \caption{Training time comparison on L40S and A100 GPU setups (5.9B model).}
    \label{fig:training-time-cpu-gpu-gpt2-5-9-b}
\end{figure}

Fig.~\ref{fig:training_time_cpu_gpu_gpt_5_9} compares the training times of \proj+P+M, StrongHold, ZeRO-Infinity, and Megatron-LM for the GPT-2~\cite{gpt4_technical_report} 5.9B model on NVIDIA L40S and NVIDIA A100 GPUs. In both settings, \proj+P+M outperforms ZeRO-Infinity and StrongHold, achieving speedups of 2.60$\times$ and 2.44$\times$ on L40S, and 2.82$\times$ and 4.17$\times$ on A100. To better understand this performance gain, Fig.~\ref{fig:one_iteration_execution_time_breakdown_gpt_5_9_b} presents a breakdown of one iteration’s execution time. The improved backward pass time of \proj+P+M is attributed to its strategic caching and prefetch-eviction mechanisms. StrongHold attains the fastest optimizer step but suffers higher forward and backward times due to frequent CPU-GPU layer transfers. Megatron-LM lacks offloading mechanism and fails to train the 5.9B model on a single GPU, resulting in a CUDA Out-Of-Memory error.


\begin{figure}[!htbp]
  \centering
\vspace{-1em}  \includegraphics[width=0.7\linewidth]{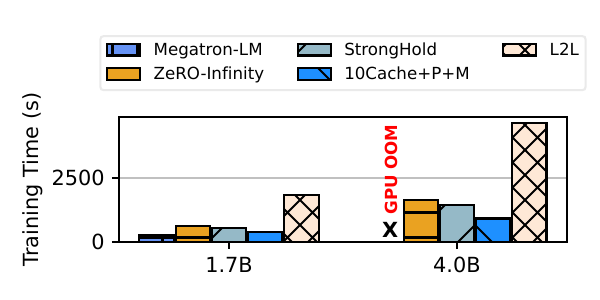}
  \vspace{-1.5em}
  \vspace{-0.5em}
  \caption{Training time comparison (1.7B and 4.0B models).
  }
  \vspace{-1em}
  \label{fig:training_time_cpu_gpu_gpt_1_7-4b}
\end{figure}

We evaluate \proj~on smaller models, GPT-2~\cite{gpt4_technical_report} 1.7B and 4.0B, using an NVIDIA L40S GPU, and compare it with Megatron-LM, L2L, StrongHold, and ZeRO-Infinity in terms of training time (Fig.~\ref{fig:training_time_cpu_gpu_gpt_1_7-4b}). \proj+P+M achieves speedups of 1.67$\times$ and 1.45$\times$ for the 1.7B model, and 1.81$\times$ and 1.58$\times$ for the 4.0B model, compared to ZeRO-Infinity and StrongHold, respectively. Megatron-LM encounters a GPU OOM error on the 4.0B model due to the lack of offloading, but performs well on the 1.7B model since all computations remain on the GPU.


\subsubsection{Wait Time Analysis (CPU-GPU Offloading)}\label{wait_time_cpu_gpu}

\begin{figure*}[ht]
    \centering
    \begin{subfigure}[b]{0.3\textwidth}
        \centering
        \includegraphics[width=\textwidth]{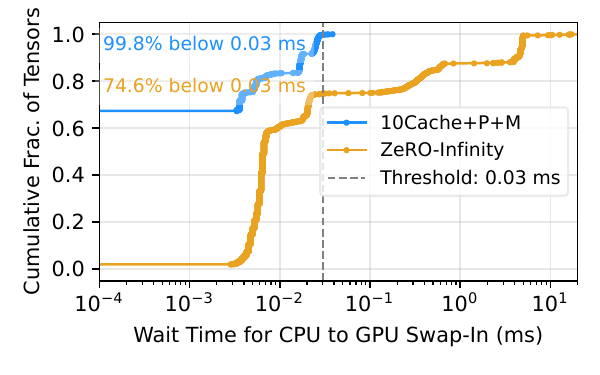}
        \vspace{-2.2em}
        \caption{OPT-6.7B}
    \end{subfigure}\hfill
    \begin{subfigure}[b]{0.3\textwidth}
        \centering
        \includegraphics[width=\textwidth]{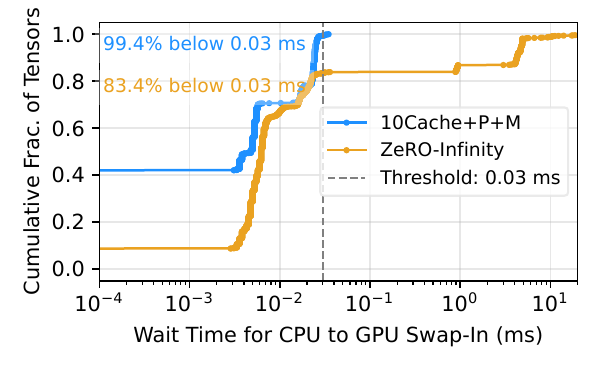}
        \vspace{-2.2em}
        \caption{Bloom-7B}
    \end{subfigure}\hfill
    \begin{subfigure}[b]{0.3\textwidth}
        \centering
        \includegraphics[width=\textwidth]{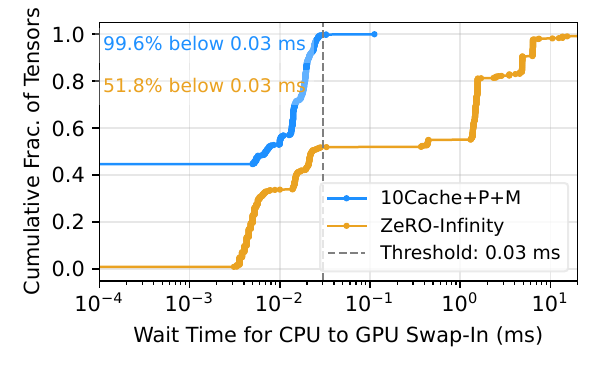}
        \vspace{-2.2em}
        \caption{Falcon-7B}
    \end{subfigure}
    \vspace{-0.6em}
    \vspace{-0.6em}
    \caption{Model parameters wait time analysis for CPU-to-GPU transfer across models (OPT-6.7B, Bloom-7B, Falcon-7B).}
    \label{fig:wait_time_cpu_gpu}
\end{figure*}

The wait time metric measures how well parameter transfers from CPU to GPU overlap with GPU computation. Fig.~\ref{fig:wait_time_cpu_gpu} compares this metric for \proj+P+M and ZeRO-Infinity, providing a fair comparison since both offload parameter tensors, unlike L2L~\cite{l2l} and StrongHold~\cite{stronghold-SC22}, which move model layers between CPU and GPU. Fig.~\ref{fig:wait_time_cpu_gpu} shows the percentage of tensors with wait times below 0.03~ms~(\cref{tensor_behavior_analysis}), computed as ((tensors’ count with wait time below 0.03 ms * 100) / total tensor count). With pre-allocated GPU cache, \proj~loads tensors before execution for direct GPU access, while its prefetching further overlaps data transfers with computation more effectively than ZeRO-Infinity. Consequently, for OPT-6.7B, Bloom-7B, and Falcon-7B, \proj+P+M achieves 1.36$\times$, 1.19$\times$, and 1.92$\times$ more tensors with wait times below 0.03~ms than the baseline.



\begin{figure}[ht]
    \centering
    \vspace{-1em}
    \begin{subfigure}[b]{0.30\textwidth}
        \centering
        \includegraphics[width=\textwidth]{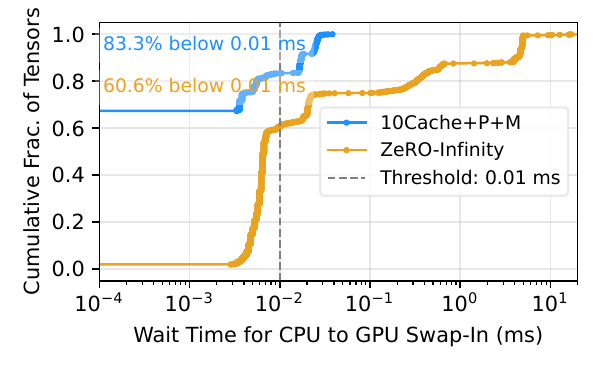}
        \vspace{-2.2em}
        \caption{Threshold 0.01ms}
    \end{subfigure}\hfill
    \begin{subfigure}[b]{0.30\textwidth}
        \centering
        \includegraphics[width=\textwidth]{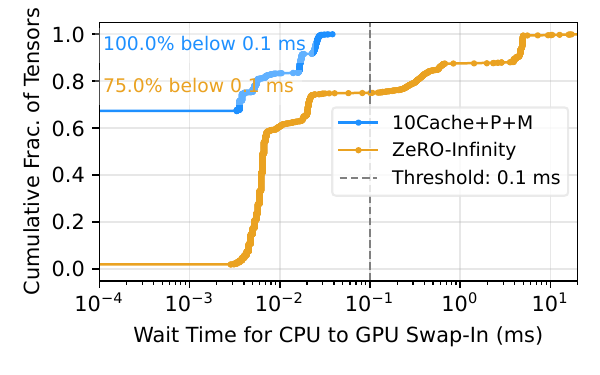}
        \vspace{-2.2em}
        \caption{Threshold 0.1ms}
    \end{subfigure}
    \vspace{-1em}
    \caption{
    OPT-6.7B parameters CPU-to-GPU transfer wait time at varying thresholds.
    }
    \label{fig:wait_time_cpu_gpu_varried_threshold}
\end{figure}

Fig.~\ref{fig:wait_time_cpu_gpu_varried_threshold} shows the proportion of tensors with CPU-to-GPU transfer wait times below 0.01~ms and 0.1~ms. With a 0.01~ms threshold, about 83\% of tensors meet the target using \proj, and at 0.1~ms, nearly all tensors fall below the threshold, highlighting \proj’s ability to minimize swap-in delays across varying tolerances.


\subsubsection{GPU Cache Hit Rate Analysis (CPU-GPU Offloading)} \label{hit_rate_cpu_gpu}

\begin{figure}
  \centering
  \vspace{-0.5em}\includegraphics[width=0.7\linewidth]{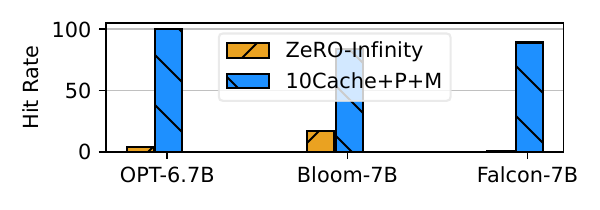}
  \vspace{-1.8em}
  \caption{
  GPU cache hit rate comparison.
  }
  \label{fig:hit_rate_cpu_gpu}
\end{figure}

Training time is closely tied to the frequency of tensor offloading. More offloading increases training time, while a higher hit rate reduces it by serving more tensors directly from GPU cache. Fig.~\ref{fig:hit_rate_cpu_gpu} compares the hit rates of \proj+P+M and ZeRO-Infinity, showing that \proj’s optimized prefetch-eviction significantly improves GPU cache efficiency. Specifically, \proj+P+M achieves 27.17$\times$, 4.87$\times$, and 86.6$\times$ higher hit rates for OPT-6.7B, Bloom-7B, and Falcon-7B, respectively, than the baseline.



\subsubsection{Training Time Evaluation (CPU-GPU-NVMe Offloading)} \label{cpu_gpu_nvme_training_time}

\begin{figure}
  \centering
  \vspace{-1em}
  \includegraphics[width=0.7\linewidth]{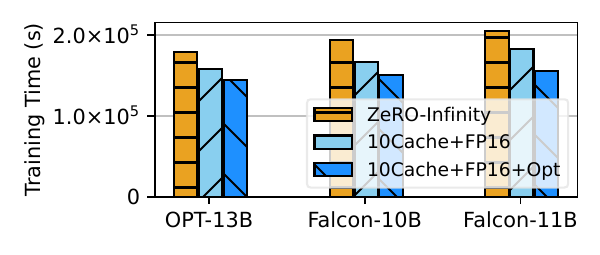}
  \vspace{-1.8em}
  \caption{
  Training performance of \proj~vs. ZeRO-Infinity under CPU-GPU-NVMe offloading.
  }
  \vspace{-1em}
  \label{fig:training_time_cpu_gpu_nvme_10b_range}
\end{figure}

Fig.~\ref{fig:training_time_cpu_gpu_nvme_10b_range} compares training time for OPT-13B, Falcon-10B, and Falcon-11B. \proj+FP16 schedules FP16 parameters while storing all optimizer states in NVMe. During the optimizer step, it sequentially loads optimizer states from NVMe to CPU memory for parameter updates. In contrast, \proj+FP16+Opt improves this by storing immediately required optimizer states in CPU memory (\cref{CPU-GPU-NVMe Offloading}) and asynchronously fetching others during updates. This reduces training time by about 20\% for OPT-13B, 23\% for Falcon-10B, and 25\% for Falcon-11B. ZeRO-Infinity suffers from performance inefficiencies due to its scheduling policy, which places all optimizer states in NVMe memory. Excess data offload 
leads to longer training times, highlighting the advantage of \proj’s optimized scheduling.

Table~\ref{tab:previous_approach_comparison} compares per-iteration training times of \proj~and baseline methods on the COLA dataset using the BERT~\cite{bert} model with a batch size of 128 on an NVIDIA A40 GPU. \proj~achieves up to a 19$\times$ speedup over prior GPU memory swapping approaches through efficient tensor caching and optimized prefetch-eviction. In the CPU–GPU–NVMe offloading setup (Table~\ref{tab:previous_approach_comparison}), \proj~outperforms ZeRO-Infinity as its multi-tiered memory bandwidth-aware tensor allocator prioritizes GPU memory, selectively spills to CPU, and minimizes NVMe use. In contrast, ZeRO-Infinity places many tensors in NVMe, increasing migration latency, while \proj’s optimized placement yields significant performance gains.

\begin{table}[h]
\centering
\vspace{-1em}
\caption{
Per-iteration training time comparison.
}
\label{tab:previous_approach_comparison}
\vspace{-1em}
\renewcommand{\arraystretch}{1}
\setlength{\tabcolsep}{0.1pt} 
\begin{tabular}{ | c | c | } 
 \hline
Approach & \makecell{
Per-Iter Time (ms)
} \\ 
 \hline
 FlashNeuron & 5905.97 \\ 
 \hline
 DeepUM & 7566.76 \\ 
 \hline
 G10 & 6667.69 \\  
 \hline
 \textcolor{blue}{\proj}~(CPU-GPU Offloading) & \textcolor{blue}{381.70} \\
 \hline
 ZeRO-Infinity (CPU-GPU Offloading) & 462.79 \\
 \hline
 \textcolor{blue}{\proj}~(CPU-GPU-NVMe Offloading) & \textcolor{blue}{401.46} \\
 \hline
 ZeRO-Infinity (CPU-GPU-NVMe Offloading) & 7611.66 \\
 \hline
\end{tabular}
\end{table}


\subsubsection{Wait Time Analysis for FP16 Parameters (CPU-GPU-NVMe)}\label{wait_time_cpu_gpu_nvme}

\begin{figure*}[ht]
    \centering
    \vspace{-0.8em}
    \begin{subfigure}[b]{0.3\textwidth}
        \centering
        \includegraphics[width=\textwidth]{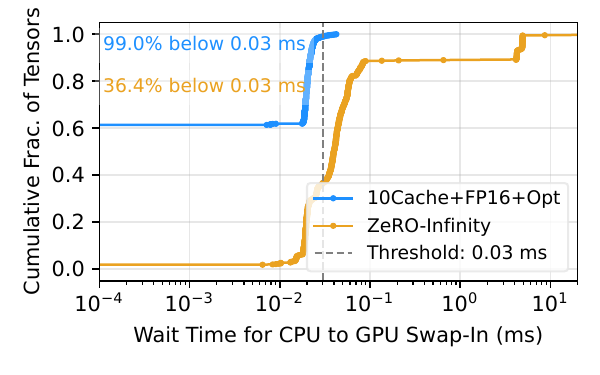}
        \vspace{-2.2em}
        \caption{Falcon-10B}
    \end{subfigure}\hfill
    \begin{subfigure}[b]{0.3\textwidth}
        \centering
        \includegraphics[width=\textwidth]{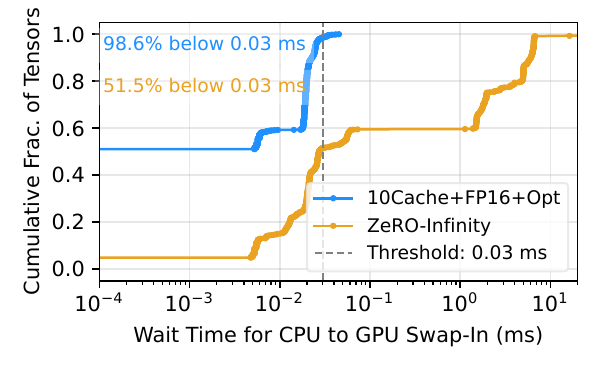}
        \vspace{-2.2em}
        \caption{Falcon-11B}
    \end{subfigure}\hfill
    \begin{subfigure}[b]{0.3\textwidth}
        \centering
        \includegraphics[width=\textwidth]{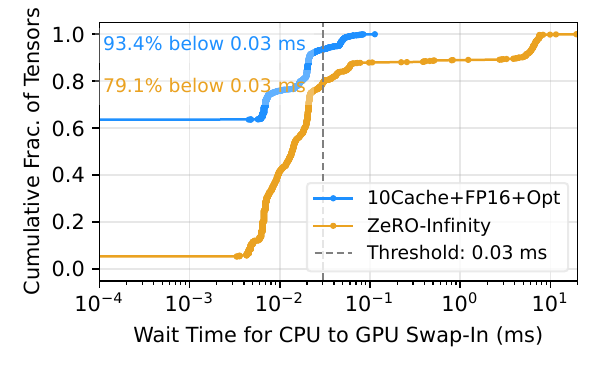}
        \vspace{-2.2em}
        \caption{OPT-13B}
    \end{subfigure}
    \vspace{-0.6em}
    \vspace{-0.6em}
    \caption{Model parameters wait time analysis for CPU-to-GPU transfer across models (Falcon-10B, Falcon-11B, OPT-13B).}
    \label{fig:wait_time_cpu_gpu_nvme_param16}
\end{figure*}

\proj+FP16+Opt processes 2.71$\times$ more tensors with wait times below 0.03~ms (\cref{tensor_behavior_analysis}) for Falcon-10B compared to ZeRO-Infinity (Fig.~\ref{fig:wait_time_cpu_gpu_nvme_param16}), with improvements of 1.91$\times$ for Falcon-11B and 1.18$\times$ for OPT-13B. Through strategic tensor placement, optimized prefetch-eviction, and efficient GPU cache use, \proj~significantly reduces data offloading overhead, leading to efficient training execution.



\subsubsection{
FP16 Parameter Count Analysis in NVMe
}

\begin{figure}[t]
  \centering
  \vspace{-0.6em}
  \includegraphics[width=0.7\linewidth]{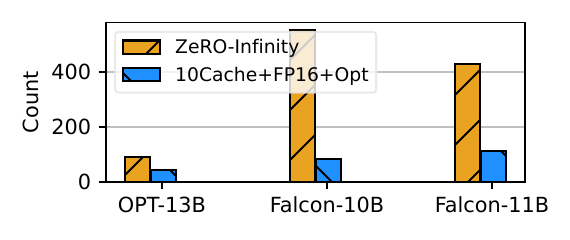}
  \vspace{-1em}
  \vspace{-0.3cm}
  \caption{
  FP16 parameter count in NVMe across models.
  }
  \label{fig:param16_nvme_count}
\end{figure}

The tensor allocator~(\cref{tensor_allocator}) in \proj~optimally places tensors across heterogeneous storage to minimize data offloading overhead. We evaluate its effectiveness by measuring the number of FP16 parameters stored in NVMe for \proj+FP16+Opt and the baseline in three models. \proj+FP16+Opt reduces the count in NVMe by 2.1$\times$ for OPT-13B, 6.7$\times$ for Falcon-10B, and 3.8$\times$ for Falcon-11B (Fig.~\ref{fig:param16_nvme_count}), demonstrating that \proj's optimized tensor placement effectively reduces data offloading and improves training efficiency.



\subsubsection{Cache Hit Rate and Miss Rate} \label{hit_rate_cpu_gpu_nvme}

\begin{figure}[!t]
\centering
\vspace{-1em}
\vspace{-0.6em}
\begin{subfigure}[b]{0.40\textwidth}
    \centering
    \includegraphics[width=0.75\textwidth]{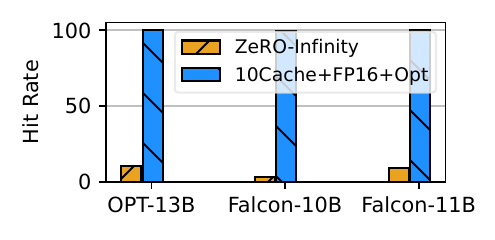}
    \vspace{-1em}
    \caption{Hit rate (FP16 parameters).}
        \label{fig:Hit-rate-for-FP16-parameter}
\end{subfigure}
\begin{subfigure}[b]{0.40\textwidth}
    \centering
    \includegraphics[width=0.75\textwidth]{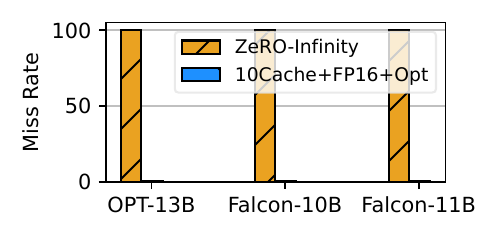}
    \vspace{-1em}
    \caption{Miss rate (FP32 optimizer states).}
    \label{fig:Miss-rate-for-FP32-optimizer-states}
\end{subfigure}
\vspace{-1.2em}
\caption{
Cache hit and miss rate comparison.
}
\vspace{-1em}
\label{fig:hit-miss-rate-param16-param32}
\end{figure}

In CPU-GPU-NVMe offloading, the FP16 parameter hit rate reflects GPU cache efficiency. Fig.~\ref{fig:Hit-rate-for-FP16-parameter} shows that \proj+FP16+Opt achieves up to 30.1$\times$ higher hit rate than ZeRO-Infinity through optimized prefetch-eviction and efficient GPU caching, while the baseline suffers from inefficient tensor scheduling.




The miss rate measures the effectiveness of \proj’s optimizer state tensor scheduler. As shown in Fig.~\ref{fig:Miss-rate-for-FP32-optimizer-states}, \proj+FP16+Opt keeps the miss rate below 1\% in all models. This efficiency comes from caching selected optimizer states in CPU memory and asynchronously loading others from NVMe when needed (\cref{CPU-GPU-NVMe Offloading}). In contrast, ZeRO-Infinity exhibits a much higher miss rate (around 100\%), as it stores all optimizer states in NVMe and retrieves them synchronously, resulting in significant performance overhead.

\subsubsection{CPU-GPU Memory Utilization} \label{cpu_gpu_memory_utilization}


We evaluate CPU and GPU memory utilization with \proj~to understand how caching strategies, including memory allocation and tensor placement, improve CPU and GPU memory efficiency compared to the baseline.



\textbf{CPU-GPU Offloading.} Both ZeRO-Infinity and \proj~store optimizer states in CPU, resulting in a small difference in CPU memory utilization. However, \proj~improves CPU usage through
pre-allocated cache memory for efficient offloading. As shown in Table~\ref{tab:cpu-gpu-offloading-mem-uti}, \proj~achieves higher GPU memory utilization by caching more tensors than ZeRO-Infinity, leading to improved computational efficiency through better GPU memory management.


\begin{table}[ht]
\centering
\vspace{-0.6em}
\caption{Memory utilization (CPU-GPU offloading).}
\label{tab:cpu-gpu-offloading-mem-uti}
\vspace{-1em}
\renewcommand{\arraystretch}{1}
\setlength{\tabcolsep}{0.1pt} 
\begin{tabular}{ | c | c | c | c | c | } 
 \hline
 Model & \makecell{\proj \\ (CPU Uti.)} & \makecell{ZeRO-Infinity \\ (CPU Uti.)} & \makecell{\proj \\ (GPU Uti.)} & \makecell{ZeRO-Infinity \\ (GPU Uti.)} \\ 
 \hline
 OPT-6.7B & 76\% & 70\% & 90\% &  76\%\\ 
 \hline
 Bloom-7B & 82\% & 72\% & 94\% & 78\% \\ 
 \hline
 Falcon-7B & 80\% & 72\% & 92\% & 78\% \\  
 \hline
\end{tabular}
\end{table}

\textbf{CPU-GPU-NVMe Offloading.} Table~\ref{tab:cpu-gpu-nvme-offloading-mem-uti} reveals a stark contrast in CPU memory utilization between the two approaches. \proj~effectively utilizes CPU memory to store more tensors, while ZeRO-Infinity relies heavily on slower NVMe storage, resulting in up to 2.15$\times$ lower CPU memory utilization. \proj~also achieves higher GPU memory utilization (up to 1.33$\times$), further highlighting its ability to cache more tensors directly in GPU. In contrast, ZeRO-Infinity’s inefficient scheduling limits memory optimization across CPU, GPU, and NVMe.




\begin{table}[h]
\centering
\vspace{-0.6em}
\caption{Memory utilization (CPU-GPU-NVMe offloading).}
\label{tab:cpu-gpu-nvme-offloading-mem-uti}
\vspace{-1em}
\renewcommand{\arraystretch}{1.0}
\setlength{\tabcolsep}{0.1pt} 
\begin{tabular}{ | c | c | c | c | c | } 
 \hline
 Model & \makecell{\proj \\ (CPU Uti.)} & \makecell{ZeRO-Infinity \\ (CPU Uti.)} & \makecell{\proj \\ (GPU Uti.)} & \makecell{ZeRO-Infinity \\ (GPU Uti.)} \\ 
 \hline
 OPT-13B & 82\% & 38\% & 90\% &  68\%\\ 
 \hline
 Falcon-10B & 76\% & 33\% & 94\% & 74\% \\ 
 \hline
 Falcon-11B & 76\% & 36\% & 88\% & 66\% \\  
 \hline
\end{tabular}
\end{table}


\subsubsection{Overhead Analysis of Profiling and Cache Memory Allocation} \label{profiling_overhead}

Table~\ref{tab:profile_time} reports the profiling and cache allocation times for different model sizes. For OPT-6.7B, \proj~incurs only about 1\% profiling time and 3.83\% cache allocation time relative to a single training epoch. These results indicate that \proj~adds minimal overhead, which becomes increasingly insignificant for larger models or longer training runs.

\begin{table}[h]
  \centering
  \vspace{-0.6em}
  \caption{Profiling and cache memory allocation time.}
  \vspace{-1em}
  \renewcommand{\arraystretch}{1}
  \setlength{\tabcolsep}{0.1pt} 
  \begin{tabular}{|l|l|l|l|}
    \hline
    & \textbf{Model Name} & \makecell{\textbf{Profile}\\\textbf{Time (s)}} & \makecell{\textbf{Pre-allocate}\\\textbf{Cache Time (s)}} \\
    \hline
    \multirow{3}{*}{CPU-GPU} & OPT-6.7B & 19.23 (1.09\%) & 67.26 (3.83\%) \\
    \cline{2-4}
    & Bloom-7B & 19.96 (0.58\%) & 71.17 (2.07\%) \\
    \cline{2-4}
    & Falcon-7B & 19.62 (0.19\%) & 70.84 (0.67\%) \\
    \hline
    \multirow{3}{*}{CPU-GPU-NVMe} & Falcon-10B & 108.54 (0.07\%) & 88.94 (0.06\%) \\
    \cline{2-4}
    & Falcon-11B & 119.12 (0.08\%) & 87.09 (0.06\%) \\
    \cline{2-4}
    & OPT-13B & 36.90 (0.03\%) & 69.83 (0.05\%) \\
    \hline
  \end{tabular}
  \label{tab:profile_time}
\end{table}

\subsubsection{Batch Size Impact on Training Time} \label{batch_size_impact_on_training_time}

\begin{figure}[!t]
  \centering
\includegraphics[width=0.7\linewidth]{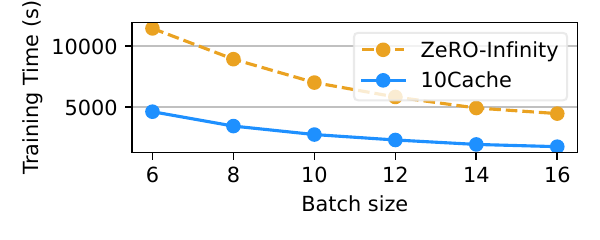}
  \vspace{-1em}
  \vspace{-5pt}
  \caption{Training time for various batch sizes (OPT-6.7B).}
  \vspace{-2em}
  \vspace{5pt}
  \label{fig:batch_size_cpu_gpu_opt_6_7_b}
\end{figure}

Fig.~\ref{fig:batch_size_cpu_gpu_opt_6_7_b} shows the OPT-6.7B model training time under CPU-GPU setup for varying batch sizes. As the batch size increases, \proj~consistently reduces training time and outperforms ZeRO-Infinity.


In summary, our results show that \proj's optimal tensor placement maximizes memory usage, achieving up to 2.15$\times$ CPU and 1.33$\times$ GPU memory utilization, significantly outperforming the baseline (\cref{cpu_gpu_memory_utilization}). Its smart caching with optimized prefetching and eviction boosts GPU cache hit rate by up to 86.6$\times$ (\cref{hit_rate_cpu_gpu}). Moreover, \proj~adds minimal profiling overhead (\cref{profiling_overhead}), ensuring efficient integration into the training workflow.

\section{Related Work}\label{related_work}



GPUs enable efficient DNN training through high computational power, but the rapid increase in model size makes GPU memory a major bottleneck~\cite{beyond_the_memory_wall}. Prior works~\cite{swapadvisor, autotm, superneuron} mitigate this by extending GPU memory with CPU-based swapping, mainly for CNNs, but performance degrades when the CPU handles memory-intensive tasks. FlashNeuron~\cite{flashneuron-FAST21} offloads intermediate tensors to SSDs to reduce CPU load, yet NVMe access latency remains a challenge. Sentinel~\cite{sentinel} dynamically profiles tensors using TensorFlow runtime and OS page faults. It maps page-level profiling to tensors by assigning each tensor a dedicated memory page and adding layer-end annotations, which greatly increase memory footprint and profiling overhead, issues that might grow with LLMs. In contrast, \proj~performs a single lightweight profiling iteration using PyTorch hooks to capture tensor execution order directly, making it more efficient and robust. While Sentinel migrates long-lived tensors based on memory access frequency, \proj~prefetches/evicts tensors by execution order, keeping needed tensors in fast memory and preventing GPU stalls. Another line of works use CUDA Unified Memory with page prefetching~\cite{deepum-ASPLOS23, g10-MICRO23}. DeepUM~\cite{deepum-ASPLOS23} profiles memory access patterns via GPU page faults to improve memory management but remains limited by CPU and GPU capacity, hindering scalability for large models. G10~\cite{g10-MICRO23} unifies CPU, GPU, and flash memory and supports page-level tensor migration. However, these methods overlook the distinct memory behavior of LLMs~(\cref{llm_memory_requirement}). \proj~addresses this by managing memory with pre-allocated cache buffers and fine-grained tensor-level migration across CPU, GPU, and NVMe based on execution order, enabling efficient training of billion-parameter LLMs.

Recent studies~\cite{zero-SC20, zero_offload-ATC21, l2l, megatron-lm, stronghold-SC22} propose memory management techniques to scale LLM training. ZeRO-Offload~\cite{zero_offload-ATC21} reduces GPU memory use by offloading gradients and optimizer states to CPU but lacks parameter offloading. ZeRO-Infinity~\cite{zero_infinity-SC21} adds NVMe offloading but its inefficient tensor placement across CPU, GPU, and NVMe leads to suboptimal memory utilization. In contrast, \proj~offers finer memory-aware placement across all tiers, improving memory utilization~(\cref{tab:cpu-gpu-offloading-mem-uti},~\cref{tab:cpu-gpu-nvme-offloading-mem-uti}) and GPU cache hit rates~(\cref{hit_rate_cpu_gpu},~\cref{hit_rate_cpu_gpu_nvme}). L2L~\cite{l2l} keeps only the active layer in GPU, incurring a high CPU-GPU communication cost, while StrongHold~\cite{stronghold-SC22} manages a sliding window of active layers in GPU that requires window size tuning. Poor tuning underutilizes GPU memory and increases layer transfer overhead. In contrast, \proj~dynamically places tensors based on memory availability, eliminating tuning and reducing data movement through efficient caching and scheduling, thereby maximizing memory efficiency and minimizing migration overhead. SHADE~\cite{khan2023shade} and FedCaSe~\cite{khan2024fedcase} apply importance-aware sampling and caching to scale deep and federated learning but focus only on in-memory caching for computer-vision workloads. In contrast, \proj~optimizes LLM fine-tuning via fine-grained tensor migration across a multi-tier cache hierarchy.


Arif et al.~\cite{cxl_optimizer_offloading} offload optimizer states to CXL memory but not parameter tensors (up to 58\% of memory, Fig.~\ref{fig:model_memory_requirement}), limiting performance in resource-constrained settings and adding communication overhead~\cite{smartinfinity-HPCA24}. MemAscend~\cite{memascend} addresses SSD-to-GPU transfer bottlenecks due to CPU buffer pool and internal fragmentation of fixed-size buffers using an adaptive pool. In contrast, \proj~addresses these issues with a smaller buffer pool and efficient memory management. Both works lack tensor lifetime analysis and CPU/GPU cache reuse, which are critical in constrained environments. While Mixture-of-Experts (MoE) and sparsity reduce memory usage, they add complexity (e.g., load balancing and communication overhead)~\cite{samoyeds, fsmoe}. \proj~focuses on dense transformers and complements sparsity by managing memory for both active and selectively used tensors. ZenFlow~\cite{zenflow} is a recent training framework that offloads GPU memory by prioritizing parameters and splitting updates between the GPU and CPU to reduce GPU stalls and I/O overhead. SuperOffload~\cite{superoffload} introduces a Superchip-centric~\cite{nvidia_superchip} system that efficiently leverages the Hopper GPU, Grace CPU, and NVLink-C2C interconnect through adaptive weight offloading, bucket repartitioning, and an optimized Adam optimizer. Both ZenFlow and SuperOffload are orthogonal to \proj~and can be integrated with it to further enhance training throughput.

\section{Discussion}\label{discussion}

\proj~significantly reduces training time for billion-scale LLMs through efficient memory management, though parts of its design still require further exploration and refinement.




\textbf{Dynamic Execution Graph.} \proj~targets LLM workloads with static execution graphs, where operation order and tensor access patterns are predictable and repeated across training iterations. This predictability enables \proj~to precompute prefetching and eviction schedules. Although dynamic execution graphs~\cite{dycl, grape, sod2, dynn_training} are gaining attention, \proj~does not yet support them due to their unpredictable nature. Future work will focus on extending \proj~to learn and adapt to dynamic execution patterns at runtime for optimized prefetching and scheduling.



\textbf{Distributed Environment.} \proj~explicitly targets the single-GPU setup, which is highly relevant for resource-constrained cloud users who cannot afford large GPU clusters. Extending \proj~to distributed training (e.g., pipeline or tensor parallelism) is a key future direction. In such settings, network communication and resource contention can make execution less predictable. To adapt, \proj’s prefetch table can be extended to record tensor execution order and GPU ownership, enabling the scheduler to prefetch and evict based on memory hierarchy and inter-GPU communication.


\textbf{Inference.} The techniques proposed in \proj~have not yet been evaluated alongside established methods such as prefill/decode optimization, KV-cache management, or workload-aware request scheduling~\cite{vllm, distserve, seesaw, flexgen, thunderserve, khan2024fairserve}. In future work, we will integrate our mechanisms, e.g., efficient tensor offloading, into a unified LLM serving architecture optimized for constrained-memory settings.

\section{Conclusion}\label{conclusion}

We present \proj, a novel framework that accelerates billion-scale LLM training for the GPU memory swapping mechanism. It features a lightweight profiler that analyzes tensor execution order to optimize placement across heterogeneous memory tiers, maximizing memory utilization, and leveraging the full memory hierarchy to boost training throughput. \proj~improves performance through efficient cache allocation, reuse of memory buffers, and intelligent prefetching and eviction that overlap data transfers with GPU computation. Experimental results show that \proj~reduces training time by about 2$\times$, increases GPU cache hit rate by up to 86.6$\times$, and achieves CPU and GPU memory utilization by 2.15$\times$ and 1.33$\times$, respectively. These results show that \proj~effectively alleviates memory bottlenecks, accelerates LLM training, and provides a practical solution for cloud environments.

\section{Acknowledgments}\label{acknowledgement}

We thank the anonymous reviewers for their valuable feedback. Some results were obtained using the Chameleon testbed~\cite{chameleon_cloud}, supported by NSF. This work is supported in part by NSF grants CSR-2106634 and CSR-2312785.


\balance
\bibliographystyle{ACM-Reference-Format}
\bibliography{references/citation}










\end{document}